\documentstyle[12pt]{article}
\def\beq{\begin{equation}}
\def\eeq{\end{equation}}
\def\bea{\begin{eqnarray}}
\def\eea{\end{eqnarray}}
\def\bq{\begin{quote}}
\def\eq{\end{quote}}
\def\PL{{ \it Phys. Lett.} }
\def\PRL{{\it Phys .Rev. Lett.} }
\def\NP{{\it Nucl. Phys.} }
\def\PR{{\it Phys. Rev.} }

\def\gappeq{\mathrel{\rlap {\raise.5ex\hbox{$>$}}
{\lower.5ex\hbox{$\sim$}}}}

\def\lappeq{\mathrel{\rlap{\raise.5ex\hbox{$<$}}
{\lower.5ex\hbox{$\sim$}}}}
\begin{document}
\topmargin -0.5cm
\oddsidemargin -0.8cm
\evensidemargin -0.8cm
\pagestyle{empty}
\begin{flushright}
CERN-TH.7214/94\\
NUB-TH.3093/94\\
CTP-TAMU-32/94\\
\end{flushright}
\vspace*{5mm}
\begin{center}
{\bf $b \rightarrow s\gamma$ DECAY IN SUPERGRAVITY GRAND UNIFICATION AND
DARK MATTER}\\
\vspace*{1cm}
{\bf Pran Nath}\\
Theoretical Physics Division,$CERN^{*}$\\
CH-1211 Geneva 23\\
\vspace*{0.5cm}
and\\
\vspace*{0.5cm}
R. Arnowitt\\
Center for Theoretical Physics, Dept. of Physics\\
Texas A\&M University, College Station, TX 77843, U.S.A.\\

\vspace*{2cm}
Abstract
\end{center}

	  An analysis of the  $b \rightarrow s\gamma$ is given in supergravity
	grand
 	unification using the framework of the radiative breaking of the
	electro-weak symmetry under three separate sets of constraints:1)
	neutralino relic density does not overclose the universe,2)p-stability
	constraint, and 3) the combined constraints of p-stability and COBE
	data .For case 1 it is found that the CLEO II data on the branching
	ratio already imposes very strong further constraints on dark matter
	analyses.
   	For case2 the
	branching ratio is found to lie in the range (1.5-6.3)x$10^{-4}$
	and thus the data does not at present significantly
	limit the analyses with p-stability constraint. It is shown that
	improvements by a factor of 3 in the  $p \rightarrow\bar\nu K^{+}$
	 lifetime will
	reduce SUSY effects to less than O(30$\%$) of the Standard Model
	value.For case3,the branching ratio lies in the
	interval (3.1-5.3)x$10^{-4}$ ,and the SUSY effects
	lie within (-10$\%$,+50$\%$) of the SM value. Thus as experimental
	bounds on $B(b \rightarrow s\gamma)$ improve,
	one would need in cases 2 and 3 the next-to-leading order QCD
	corrections to disentangle SUSY effects.

\vspace*{1.0cm}
\noindent
\rule[.1in] {17.0cm}{.002in}
\noindent
$^{*)}$ Permanent address:  Department of Physics, Northeastern
University, Boston,
MA 02115, U.S.A. \vspace*{2.0cm}
\begin{flushleft}
CERN-TH.7214/94\\
NUB-TH.3093/94\\
CTP-TAMU-32/94\\
March 1994
\end{flushleft}
\vfill\eject
\pagestyle{empty}
\clearpage\mbox{}\clearpage
\setcounter{page}{1}
\pagestyle{plain}

Last year the CLEO Collaboration obtained a new bound on the flavour
changing
decay mode $b \rightarrow s\gamma$ \cite{aaa}.  For the branching ratio
$B(b
\rightarrow s\gamma)$ they find the bound $B(b \rightarrow s\gamma) <
5.4 \times
10^{-4}$ at 95\% CL.  At the same time a definite observation of the
exclusive mode
$B \rightarrow K^*\gamma$ was made with a branching ratio $B(B
\rightarrow
K^*\gamma) = (4.5 \pm 1.5 \pm 0.9)\times 10^{-5}$\cite{aaa}.  These
branching ratio
measurements are expected to improve in the future, and should provide
stringent
limits on the Standard Model (SM) prediction.  In the SM, $b \rightarrow
s\gamma$
decay is induced at the one-loop level by a penguin diagram which
involves the
exchange of a $W$ boson and a $t$-quark.  If a definite discrepancy
between the SM
prediction and the experimental value is found, it would provide a
window to physics
beyond the SM.  For example, in supersymmetry there are additional
penguin diagrams
involving the exchange of the charged Higgs, the charginos, the gluino
and the
neutralinos which contribute to the $b \rightarrow s\gamma$ decay
\cite{bb}.  It is
already known that significant contributions can arise from the charged
Higgs
\cite{cc}--\cite{ee}.  In this paper we give a detailed analysis of $b
\rightarrow
s\gamma$ branching ratio within the framework of radiative breaking of
the $E-W$ symmetry for the minimal SU(5)
supergravity
model.  The analysis is given under three different sets of constraints:
(i)
cosmological constraint on SUSY dark matter, (ii) proton stability
constraint,
and (iii) simultaneous imposition of cosmological and proton stability
constraints.

The basic formula for the branching ratio $B(b \rightarrow s\gamma)$
including the
$W$, charged Higgs and the sparticle exchange contributions to leading
order QCD
corrections is given by\cite{bb},\cite{dd},\cite{ff},\cite{ggg}
\beq
\frac{B(b \rightarrow s\gamma)}{B(b \rightarrow ce\nu)} =
\frac{6\alpha}{\pi}~
\frac{ \left[\eta^{\frac{16}{32}}A_\gamma +
\frac{8}{3}\left(\eta^{\frac{14}{23}} -
\eta^{\frac{16}{32}}A_g\right)+C\right]^2}
{P\left(\frac{m_c}{m_b}\right)\left[1 -
\frac{2}{3\pi}\alpha_s(m_b)f\left(\frac{m_c}{m_b} \right) \right]}
\label{1}
\eeq
where $\eta = \alpha_s(m_Z)/\alpha_s(m_b)$, $P$ is a phase-space factor
defined by
$P(x) = 1-8x^2 + 8x^6-x^8 - 24x^4\ln x$ and $f(m_c/m_b)$ is a QCD
correction factor
to the process $b \rightarrow ce\nu$ for which we use the value
$f(m_c/m_b) =
2.41$.  For $B(b \rightarrow ce\nu)$ we use the experimental value 0.107
and
$A_{\gamma, g}$ are the contributions from penguin diagrams at scale
$M_W$ with
photonic or gluonic external legs but including the exchange of $W$,
charged Higgs
and the sparticle exchanges.  For $C$, the operator mixing coefficient,
we use the
valuation of Ref.~\cite{hh}:
\beq
C = \sum^8_{i=1} q_i\eta^{p_i}
\label{2}
\eeq
where $q_i$ (obeying $\sum^8_{i=1} q_i = 0)$ and $p_i$ are given by
$\{q_i;i=1,...,8 \} = (2.2996, -1.088, -0.4286,$ $-0.0714, -0.6494,
-0.038, -0.186,
-0.0057);$  $\{p_i;i=1,...,8\} = (0.6087, 0.6957, 0.2609,\linebreak
-0.5217, 0.4086,
-0.423, -0.8994, 0.1456)$.  The $(q_i,p_i)$ of Eq.~(3) are computed
using an $O(g^2)$
calculation of an $8 \times 8$ anomalous dimension matrix which enters
in the
evolution of the current-current, QCD and ``magnetic penguin" operators
as one
uses the renormalization group to evolve from the weak scale $M_W$ to
the scale
$\mu = O(m_b)$.  (There is a disagreement in the literature on the
numerical
evaluation of $(q_i,p_i)$ (see papers of Ref.\cite{jj}) but this
disagreement
only leads to O(1\%) differences and thus is not of measurable
significance at
this stage).

The evaluation of $B(b \rightarrow s\gamma)$ from Eq.~(\ref{1}) suffers
in general
from several uncertainties.  There are uncertainties generated due to
experimental
errors in the quark masses and $\alpha_s$.  However, potentially the
largest error
is due to the renormalization point $\mu$ dependence of Eq.~(\ref{1}).
For example
a variation of $\mu$ by a factor of 2 in each direction from its mean
value $m_b$
(i.e. in the range $m_b/2$ to $2m_b$) can generate a $\pm 25\%$
variation, in the
branching ratio \cite{kk},\cite{lll}.  	This $\mu$-dependence is
expected to be
diluted by a significant amount \cite{lll} by inclusion of the
next-to-leading order
QCD corrections, analogous to what has been observed in other FCNC
processes
\cite{mm}.  Presently, only partial analyses of the next-to-leading
order QCD
corrections exist, and the full analysis appears very involved requiring
analysis of
three loop mixings in some sectors \cite{lll}.  These next-to-leading
order
corrections are likely to become more significant as experimental
measurements of
$B(b \rightarrow s\gamma)$ improve.

In this paper we analyze Eq.~(\ref{1}) within the framework of the
supergravity
unified models \cite{nn}, \cite{oo} with an SU(5) type embedding.  In
this model
supersymmetry is broken spontaneously via a hidden sector, and the
effective
potential contains the following supersymmetry breaking terms below the
GUT scales
\cite{nn}--\cite{rr}:  $m_0, m_{1/2}, A_0$ and $B_0$, where $m_0$ is the
universal
scalar mass, $m_{1/2}$ is the universal gaugino mass and $A_0$ and
$B_0$, are the cubic and quadratic soft SUSY breaking constants
that appear
in the effective theory below the GUT scale.  Further, one finds for the
effective
superpotential of the theory the form $W_{eff} = W^{(2)} + W^{(3)} +
M^{-1}_{H_3}W^{(4)}$, where $W^{(2)} = \mu_0H_2H_1$, and $H_2(H_1)$ are
the Higgs
doublets that give mass to the up (down) quarks, $W^{(3)}$ contains the
usual cubic
interactions of the quarks (and of the leptons) to the Higgs, and
$W^{(4)}$ contains
baryon number violating dimension four operators which are responsible
for
proton decay via dimension five operators.  The $W^{(4)}$ interactions are
suppressed
by the superheavy Higgsino triplet mass $M_{H_3}$.  We use the
renormalization group
analysis to break the electroweak symmetry \cite{ss}, and after
radiative breaking
of the symmetry one can determine the parameter $\mu_0$ by fixing the
$Z$-mass, and
determine $B_0$ in terms of $\tan\beta = \langle H_2\rangle /\langle H_1
\rangle$.
The model is then completely specified by four parameters
\beq
m_0,~m_{1/2},~ A_t, \tan\beta
\label{4}
\eeq
and the sign of $\mu$.  Here $A_t$ is the value of $A_0$ at the
electroweak scale.
There are 32 new particles in this model (12 squarks, 9 sleptons, 2
charginos, 4
neutralinos, 1 gluino, 2CP even neutral Higgs, 1 CP odd neutral Higgs
and 1 charged
Higgs).  These 32 new particles and all their interactions, can be
determined in
terms of the four parameters of Eq.~(\ref{4}).  Thus the theory makes many
predictions.  A
general analysis of $b \rightarrow s\gamma$ in supergravity unification
is given in
Ref.~\cite{tt}.  Here we discuss $b \rightarrow s\gamma$ decay under
constraints of
cosmology and proton stability.

An interesting result of supergravity unification with $R$-invariance is
that over
much of the parameter space the lightest neutralino is also the lightest
supersymmetric particle (LSP) and is a natural candidate for dark
matter
\cite{uu}--\cite{yy}.  Thus at the very least one must impose the
constraint that the
neutralino dark matter not overclose the Universe, i.e., one has
\beq
\Omega_{\tilde Z_1}~h^2 < 1~.
\label{4a}
\eeq
Here $\Omega_{\tilde Z_1} = \rho_{\tilde Z_1}/\rho_c$ where
$\rho_{\tilde Z_1}$ is
the neutralino mass density, $\rho_c$ is the critical mass density  needed to
close the Universe, and $h$ is the
Hubble
constant in units of 100~km/sec~Mpc. Current measurements  give h the range
$0.5 < h < 0.75$.  Under
additional assumptions one can obtain a more stringent bound than
(\ref{4a}).  Thus
assuming the inflationary scenario which requires $\Omega = 1$, and a mix of
cold dark
matter (CDM) and hot dark matter (HDM) in the ratio 2:1 which is
consistent with COBE
data, one finds $0.1 < \Omega_{\tilde Z_1} h^2 < 0.35$ where we have
used the estimate
$\Omega_{NB} \approx 0.9$ for non-baryonic matter.  In our computation
of the
neutralino relic density we have used the accurate method which
integrates over the
Higgs and the $Z$ poles in the neutralino annihilation [22-24].

Next we discuss the $p$-stability constraint.  In the minimal model
the most dominant decay mode of the nucleon is $p \rightarrow \bar\nu
K^+(n \rightarrow
\bar\nu K^0)$ which proceeds via dimension five operators
generated from the
exchange of the Higgsino triplet field, i.e., the $W^{(4)}$ term in
$W_{eff}$ discussed
earlier.  The proton decay lifetime for this mode is \cite{aai},\cite{bbi}.
\beq
 \tau(p
\rightarrow \bar\nu K^+) = {\rm Const} \left( \frac{M_{H_3}}{\beta_p}
\right)^2
|B|^{-2}
 \label{5}
\eeq
 where the front factor is determined using chiral Lagrangian
technique, $\beta_p$ is the three-quark matrix element of the proton
which is determined
via lattice gauge theory computations, and $B$ is the dressing loop
function which
depends on masses of the SUSY particles that enter in the dressing loop.
A reasonable
upper bound on $M_{H_3}$ without the Planck scale effects becoming
dominant is $M_{H_3}
< 10~M_G$ \cite{ww} which replaces the more stringent condition of
$M_{H_3} < 3M_G$
imposed in previous analyses \cite{uu},\cite{vv},\cite{eei}.
The most recent
lattice gauge determination of $\beta_p$ gives $\beta_p = (5.6 \pm 0.8)
\times
10^{-3}~{\rm GeV}^{-3}$ \cite{cci}.

We can now use Eq.~(\ref{5}) to obtain an upper bound on $(B)$ using the
experimental lower bound on $\tau(p \rightarrow \bar\nu K^+)$ of $1.0
\times
10^{32}$~yr \cite{ddi}.  One finds that $B < 100(M_H_3
/M_G) GeV^{-1}$. The allowed range of  $p \rightarrow \bar\nu
K^+$ lifetime then is

\beq
 \tau(p
\rightarrow \bar\nu K^+) \gappeq 10^{32} \left( \frac{M_{H_3}}{M_G}
\right)^2 \left(\frac{100~{\rm GeV}^{-1}}{B}\right)^2 yr
 \label{6}
\eeq

We proceed as follows:  we use renormalization group
evolution
of gauge, Yukawa and soft SUSY breaking terms with supergravity boundary
conditions
imposed at the GUT scale $M_G$.  After the breaking of electroweak
symmetry, the
SUSY spectrum  is computed in terms of the allowed values of the parameter
space
defined in
Eq.~(\ref{4}).  As discussed in Refs.~\cite{vv},\cite{eei}, it is found
that over
much of the parameter space $|\mu| \gg M_Z$ and this leads to certain
scaling relations
on the mass spectra.  For example, for the chargino and neutralino masses
one finds

$$
m_{\tilde W_1} \simeq m_{\tilde Z_2} \simeq 2m_{\tilde Z_1}
\eqno{(7a)}
$$
$$
m_{\tilde W_1} \simeq \frac{1}{3} m_{\tilde g}~(\mu < 0)~;~~m_{\tilde
W_1} \simeq
\frac{1}{4}m_{\tilde g}~(\mu > 0)
\eqno{(7b)}
$$
Also one finds in the same domain of the parameter space that the
heavier CP even
Higgs, the CP odd Higgs and the charged Higgs are essentially degenerate
in mass,
i.e.,
\addtocounter{equation}{1}
\beq
m_{H^0} \simeq m_A \simeq m_{H^+}
\label{8}
\eeq
The analysis of Eq.~(\ref{1}) is carried out using the spectrum that
emerges from
the radiative electroweak breaking.  In the evaluation of $B(b
\rightarrow s\gamma)$
we shall include the contributions of $W, H^\pm$ and $\tilde W_i
(i=1,2)$ and
neglect the (small) contributions from the gluino and neutralino states.
The
chargino contributions involve the exchange of all the three generation
of squark
states.  Of specific interest are the exchanges of the third generation
of squarks
(i.e. stops) which can make large contributions over certain regions of
the parameter
space.   To fix notation the stop-(mass)$^2$ matrix is given by
\beq
\left(
\matrix{
m^2_{\tilde t_L} & m_t(A_t + \mu \cot \beta) \cr
m_t(A_t + \mu\cot \beta) & m_{\tilde t_R}^2 } \right)
\label{9}
\eeq
where
$$
m_{\tilde t_L}^2 = m^2_Q + m^2_t +
\left[ \left( -\frac{1}{2} \right) + \left( \frac{2}{3} \right)
\sin^2\theta_W
\right] M^2_Z \cos2 \beta
\eqno{(10a)}
$$
$$
m^2_{\tilde t_R} = m^2_U + m^2_t - \left( \frac{2}{3} \right)
\sin^2\theta_W M^2_Z
\cos2\beta
\eqno{(10b)}
$$
and $m^2_Q, m^2_U$ are as defined in Iba$\tilde{\rm n}$ez et al.
Ref.~\cite{ss}.

We find in general that for the supersymmetric case there are
significant regions of
the parameter space where $B(b \rightarrow s\gamma)$ can be either
larger or smaller
than the corresponding SM value.  An
interesting phenomenon for some specific points in the parameter space
is the almost
perfect cancellation of the branching ratio because of contributions
from the
chargino exchange.  Although supersymmetry is certainly at the root of
these almost
perfect cancellations, the points in the parameter space where such
cancellations
occur are far from the exact supersymmetric limit where one expects a
perfect
cancellation \cite{ffi}.

We discuss next details of the analysis.  As expected the charged Higgs
makes a positive
contribution and increases the $b \rightarrow s\gamma$ branching ratio.
However, the
contribution of the chargino can be either positive or negative
depending on the point in the parameter space and the
sign of $\mu$.  This conclusion agrees with previous analyses of
Refs.~\cite{dd},
\cite{ee}, and \cite{ggi} that chargino contributions can be
either
constructive or destructive.  In general the domain where the
cosmological constraint
of Eq.(\ref{4a}) is satisfied is quite large (see Figs.~1a and 2a),
while the domain
where the $p$-stability constraint of Eq.(6), is satisfied is
generally smaller
(see Figs. 1b and 2b).  The stringent constraint of $p$-stability arises
in part due
to the fact that $p$-stability limits $\tan \beta$ so that $\tan \beta
\leq 10$.  The
parameter space where constraints of eqs. (4) and (6)  are
simultaneously
satisfied is the smallest (see Figs. 1c, 1d, 2c and 2d).

 	 The recent experimental measurements suggest a
	top mass value of 174$\pm$16 GeV \cite{hhi}. At the same time
	precision electro-weak data predicts $m_t$=162$\pm$9 GeV \cite{jji}.
	We have carried out an analysis over the full top mass range
	consistent with the above analyses. In the following we discuss
	specifically the case when the running top mass $M_Z$ =160 GeV.
	Since the running top mass at $M_Z$ is about 5 $\%$ lower than
	the physical top mass \cite{kki}, this case corresponds to a physical
	top mass of about 168 GeV. The analysis is carried out
	over the full parameter space  including both signs on the value
	of $\mu$. The result of this analysis is then found
	to be valid within $10\%$ for other top masses in the range consistent
	with the CDF \cite{hhi}, LEP and SLC \cite{jji} data.

	   The result for the case $\mu<0$ is exhibited in figs
	1a-1d, and for $\mu>0$ is exhibited in Figs 2a-2d. Fig1a gives the
	branching ratio  as a funtion of $\Omega_{\tilde Z_1}h^2$ for $\mu<0$
	in the interval consistent with the cosmological constraint of
	eq(4) but without imposition of p-stability constraint of eq(5).Since
	the branching ratio is plotted as a function of
	$\Omega_{\tilde Z_1}h^2$  it is
	straightforward to further impose the COBE constraint by limiting
	the value of $\Omega_{\tilde Z_1}h^2$
	 to the range (0.1-0.35). Fig1a shows that
	cosmological constraint with or without the COBE constraint does
	not limit the current experiment. However, the current experiment
	significantly constrains the
	dark matter analysis, since the analysis with dark matter constraints
	alone allow for $B(b \rightarrow s\gamma)$  branching ratios which
	can exceed
	the current
	experimental bounds by a significant amount. Fig1b gives the
	branching ratio as a function
	of the  $p \rightarrow \bar\nu K^{+}$  lifetime consistent with the
	current experimental
	limits on this decay[28] but without the imposition of the cosmological
	consraint of eq(4). One finds that the branching ratio lies in the
	range (1.5-5.7)
	x $10^{-4}$.
        Thus the current experimental limits do not significantly constrain
	the model when the p-stability constraint is included. Fig1c is identical
	to Fig1b except that it also includes the cosmological constraint
	of eq(4).Fig1d exhibits the branching ratio as a function of
	$\Omega_{\tilde Z_1}h^2$ with inclusion of p-stabilty constraint.
	From Fig1d one
	finds that the branching ratio in the domain consistent with
	p-stability and COBE data lies in the range (3.1-5.0)x $10^{-4}$.
	An analysis similar to that for Figs1a-1d for $\mu>0$ is carried out
	for Figs2a-2d.
	From Fig2b one finds that the branching ratio with p-stabilty
	constraint lies in the interval (2.6-5.3)x$10^{-4}$, and from
	fig1d one finds that the branching ratio under the combined
	constraint of p-stability and COBE data lies in the interval
	(3.1-5.3)x $10^{-4}$. Together from Fig1 and Fig2 we find that
	with p-stability constaint alone the branching ratio lies in
	the interval (1.5-5.7)x$10^{-4}$, and  with the combined
	constraints of p-stability and COBE data the branching ratio lies
	in the interval (3.1-5.3)x $10^{-4}$. To see the variation with
	the top mass we give below the ranges of the branching ratios
	for $m_t(M_Z)=166$ GeV which correspond to the physical top mass of
	174 GeV. Here with p-stability constraint the branching ratio lies
	in the interval (2.2-6.3) x $10^{-4}$, and with the combined
	constraints of p-stability and COBE data the range of the branching
	ratio is (3.5-5.0)x $10^{-4}$. Thus the variation in the predicted
	range of the branching ratio in each case is within $10\%$, and
	thus the ranges are found to be insensitive to small changes of
	$m_t$. For comparison the Standard Model branching ratio in this
	domain is about 3.5x$10^{-4}$.\\
 	Finally, we discuss the constraints on the parameter space that
	emerge from imposition of the CLEO II bounds. For the dark matter
	analysis of case1, the current experimental limits put very
	strong constraints for the case $\mu>0$. Here one finds that if the
	top mass lies in the range 165-175 GeV, then 60-70$\%$ of
	the parameter space allowed by the constraint that neutralino relic
	density not overclose the universe, is eliminated. For the case
	$\mu<0$,
	the fraction of the parameter space consistent with the cosmological
	constraint and in violation of the CLEO II bound is much smaller,
	i.e., 15-20$\%$ of the parameter space consistent with Eq(4) is
	eliminated. The constraints on the parameter space from CLEO II
	bounds in cases 2 and 3 are negligible.

\midskip

\noindent
CONCLUSIONS\\

    We have analysed the inclusive  $B(b \rightarrow s\gamma)$ decay branching
	ratio under three different sets of constraints: cosmological
	constraint, p-stability constraint, and the combined constraints
	of p-stability and COBE data. An important result that emerges
	from the analysis is that the current experimental limits on the
  $B(b \rightarrow s\gamma)$ branching ratio put significant constraints on
	dark matter analyses for $\mu>0$. The constraints on the parameter
	space would become even more severe as the experimental bounds
	on the branching ratio improve. The theoretical analysis under the
	p-stability constraint
	gives a branching ratio in the range (1.5-6.3)x$10^{-4}$
	Thus the current CLEO bounds do not significantly constrain
	the model in this case. The analysis under the combined constraints
 	of p-stabilty and COBE data give a branching ratio in the range
	(3.1-5.3)x$10^{-4}$. Thus the result of the analysis with the
	combined constraints of p-stability and COBE data is totally
	unconstrained by current experiment.The analysis with p-stability
	also shows (see Figs 1b and 2b ) that when lifetime measurements
	on  $p \rightarrow\bar\nu K^{+}$ decay mode improve by a factor of
	about 3, the allowed
	theoretical variations in the branching ratio will fall within
	O(30$\%$) of the Standard Model prediction, and consequently the
	next-to-leading order effects will
	become relevant. However, the analysis under the constraints
	of p-stability and COBE data gives a result for the branching
	ratio at the level of the leading order calculation, which has
	a variation of O(-10$\%$,+50$\%$) from the SM value
	and a variation of only O(30$\%$) around its mean  .Thus the
	next-to-leading order effects are already relevant in this case
	if one wants to disentangle the SUSY effects when experimental
	measurements improve.

\noindent
ACKNOWLEDGEMENTS

This research was supported in part by NSF grant numbers PHY-19306906
and PHY-916593.

\vfill\eject

\vfill\eject

\noindent
{\bf FIGURE CAPTIONS}
\begin{itemize}
\item[Fig. 1a:]  Plot of the branching ratio $BR (b \rightarrow
s\gamma)$ as a
function of $\Omega_{\tilde Z_1}h^2$ for the domain
$\Omega_{\tilde Z_1}h^2 < 1$ when no
$p$-stability
constraint is imposed.  The analysis is for $m_t(M_Z)$ = 160~GeV and
$\mu < 0$.  All
other parameters are integrated out.
\item[Fig. 1b:] $BR (b \rightarrow s\gamma)$ vs. the proton lifetime
$\tau (p
\rightarrow \bar\nu K)$ without imposition of dark matter constraints.
All other
parameters are the same as in Fig. 1a.
\item[Fig. 1c:] Same as Fig. 1b except the cosmological constraint
$\Omega_{\tilde Z_1}h^2 <
1$ is imposed.
\item[Fig. 1d:] Same as Fig. 1a except that $p$-stability constraint is
imposed.
\item[Fig. 2a]:  Same as Fig. 1a except $\mu >
0$.
\item[Fig. 2b:]  Same as Fig. 1b except $\mu >
0$.
\item[Fig. 2c:]  Same as Fig. 1c except $\mu >
0$.
\item[Fig. 2d:] Same as Fig. 1d except  $\mu >
0$.
\end{itemize}


\begin{thebibliography}{99}
\bibitem{aaa} R. Ammar et al., \PRL {\bf 71} (1993) 674;\\
E. Thorndike et al., {\it Bull. Am. Phys. Soc} {\bf 38} (1993) 992.
\bibitem{bb} S. Bertolini, F. Borzumati and A. Masiero, \NP {\bf B294}
(1987) 321;\\
S. Bertolini, F. Borzumati, A. Maseiro and G. Ridolfi, \NP {\bf B353}
(1991) 591.
\bibitem{cc} J. Hewett, \PRL {\bf 70} (1993) 1045;\\
V. Barger, M. Berger and R.J.N. Phillips, \PRL {\bf 70} (1993) 1368;\\
M.A. Diaz, \PL {\bf B304} (1993) 279.
\bibitem{dd} R. Barbieri and G. Giudice, \PL {\bf B309} (1993) 86.
\bibitem{ee} N. Oshima, \NP {\bf B304} (1993) 20;\\
R. Garisto and J.N. Ng, \PL {\bf B315} (1993) 372.
\bibitem{ff} T. Inami and C.S. Lim, {\it Prog. Theor. Phys.} {\bf 65}
(1981) 297;
{\bf 65} (1981) 17772E.
\bibitem{ggg} B. Grinstein, R. Springer and M. Wise, \NP {\bf B339}
(1990) 269;\\
M. Misiak, \PL {\bf B269} (1991) 161.
\bibitem{hh} P. Cho and M. Misiak, Caltech preprint CALT-68-1893 (1993).
\bibitem{jj} M. Misiak, \NP {\bf B393} (1993) 23;\\
M. Ciuchini, E. Franco, G. Martinelli, L. Reina and L. Silvestrini, \PL
{\bf B316}
(1993) 127;\\
K. Adel and Y.P. Yao, {\it Modern. Phys. Lett} {\bf A8} (1993) 1679.
\bibitem{kk} A. Ali, C. Greub and T. Mannel, Proc. ECFA Workshop on
$B$-meson
Factory, eds. R. Aleksan and A. Ali, DESY (1993).
\bibitem{lll} A.J. Buras, M. Misiak, M. M\"unz and S. Pokorski,
Max-Planck Institute
preprint MPI-Ph/93-77 (1993).
\bibitem{mm} A.J. Buras, M. Jamin and P.H. Weisz, \NP {\bf B347} (1991)
491;\\
G. Buchella and A.J. Buras, \NP {\bf B398} (1993) 285.
\bibitem{nn} A.H. Chamseddine, R. Arnowitt and P. Nath, \PRL {\bf 29}
(1982) 970.
\bibitem{oo} P. Nath, R. Arnowitt and A.H. Chamseddine, {\it ``Applied
$N=1$
Supergravity"} (World Scientific, Singapore, 1984).
\bibitem{pp} R. Barbieri, S. Ferrara and C.A. Savoy, \PL {\bf B119}
(1983) 343.
\bibitem{qq} L. Hall, J. Lykken and S. Weinberg, \PR {\bf D22} (1983)
2359.
\bibitem{rr} P. Nath, R. Arnowitt and A.H. Chamseddine, \NP {\bf B227}
(1983) 121;\\
S.Soni and A. Weldon, \PL {\bf B216} (1983) 215.
\bibitem{ss} K. Inoue et al., {\it Prog. Theor. Phys.} {\bf 68} (1982)
927;\\
L. Iba$\tilde{\rm n}$ez and G.G. Ross, \PL {\bf B110} (1982) 227;\\
J. Ellis, J. Hagelin, D.V. Nanopoulos and K. Tamvakis, \PL {\bf 125B}
(1983) 275;\\
L. Alvarez-Gaum\'e, J. Polchinski and M.B. Wise, \NP {\bf B250} *1983)
495;\\
L.E. Iba$\tilde{\rm n}$ez, C. Lopez and C. Munos, \NP {\bf B256} (1985)
218.
\bibitem{tt} J. Wu, R. Arnowitt and P. Nath, CERN-TH.7316,CTP-TAMU-03/94,
NUB-TH.3092/94.
\bibitem{uu} S. Kelley, J. Lopez, D.V. Nanopoulos, H. Pois and K. Yuan,
\PL {\bf
B272} (1991) 423.
 \bibitem{vv} R. Arnowitt and P. Nath, \PRL {\bf 69} (1992) 725.
\bibitem{ww} R. Arnowitt and P. Nath, \PL {\bf B299} (1993) 58 and
Erratum ibid.
{\bf B303} (1993) 403.
\bibitem{yy}P. Nath and R. Arnowitt, \PRL {\bf 70} (1993) 3696.
\bibitem{zz} S. Kelley and J.L. Lopez, D.V. Nanopoulos and K. Yuan, \PR {\bf
D47}(1993) 2461.
\bibitem{aai} J. Ellis, D.V. Nanopoulos and S. Rudaz, \NP {\bf B202}
(1982) 43;\\
R. Arnowitt, A.H. Chamseddine and P. Nath, \PL {\bf 156B} (1985) 215;\\
P. Nath, R. Arnowitt and A.H. Chamseddine, \PR {\bf 32D} (1985) 2348;\\
J. Hisano, H. Murayama and T. Yanagida, \NP {\bf B402} (1993) 46.
\bibitem{bbi} R.Arnowitt and P.Nath,  \PR {\bf 49D} (1994)1479.
 \bibitem{cci} M.B. Gavela et al., \NP {\bf B312} (1989) 269.
\bibitem{ddi} R. Becker-Szendy et al., \PR {\bf D47} (1993) 4028.
\bibitem{eei} P.Nath and R.Arnowitt,\PL {\bf 289B} (1992)368.
\bibitem{ffi} S. Ferrara and E.Remiddi, \PL {\bf B53} (1974) 347.

\bibitem{ggi} J.L. Lopez, D.V. Nanopoulos and G. Park, \PR {\bf D48}
(1993) R974;\\
G.L. Kane, C. Kolda, L. Roszkowski and J.D. Wells, Univ.
Michigan
preprint UM-TH-93-24 (1993);\\
S. Bertolini and F. Vissani, SISSA Preprint, SISSA 40/94/EP.
\bibitem{hhi} CDF Collaboration,FNAL Preprint Fermi LAB-PUB-94/097-E.
\bibitem{jji} J.Ellis,G.L.Fogli and E. Lisi, CERN-TH.7261/94.
\bibitem{kki} N.Gray et al., Z.Phys.{\bf C48} (1990)673;
H. Arason et al., \PR {\bf D46} (1992) 3945.
\end{thebibliography}
\end{document}